\documentclass[10pt,a4paper]{iopart}
\pdfoutput=1
\usepackage{textcomp}
\usepackage{lmodern}     
\usepackage{iopams}

\expandafter\let\csname equation*\endcsname\relax
\expandafter\let\csname endequation*\endcsname\relax
\usepackage{amsmath}
\usepackage{amssymb}
\usepackage{amsthm}
\usepackage{amsfonts}
\usepackage{graphicx}
\usepackage{color}
\usepackage{bbm}
\usepackage{float}
\usepackage{dsfont}
\usepackage{url}
\usepackage[colorlinks=true,linkcolor=blue, citecolor=blue, urlcolor=blue, bookmarks]{hyperref}
\usepackage{subfigure}

\newcommand{\be}{\begin{equation}}
\newcommand{\ee}{\end{equation}}
\newcommand{\bea}{\begin{eqnarray}}
\newcommand{\eea}{\end{eqnarray}}
\def\nn{\nonumber\\}
\def\doi{http://dx.doi.org/}
\def\fr#1{(\ref{#1})}

\def\rrangle{\rangle\!\rangle} 
\def\llangle{\langle\!\langle}
\def\Bigrrangle{\Big\rangle\!\Big\rangle} 
\def\Bigllangle{\Big\langle\!\Big\langle}

\def\ontop#1#2{\genfrac{}{}{0pt}{}{#1}{#2}}
\def\sfix#1{\texorpdfstring{#1}{Lg}}
\usepackage{bbm} 
\usepackage{graphicx}

\def\id{\mathbbm{1}}
\def\idc{\mathbbm{1}_{\rm cl}}

\usepackage{color}

\begin{document}
\title[Exact solution of a quantum ASEP with particle creation and annihilation]{Exact solution of a quantum asymmetric exclusion process with
 particle creation and annihilation} 

\author{Jacob Robertson and Fabian H L Essler}

 \address{The Rudolf Peierls Centre for Theoretical Physics, Oxford
  University, Oxford OX1 3PU, United Kingdom}
  \ead{fab@thphys.ox.ac.uk}
\date{\today}
\begin{abstract}
We consider a Lindblad equation that for particular initial conditions
reduces to an asymmetric simple exclusion process with additional
loss and gain terms. The resulting Lindbladian exhibits
operator-space fragmentation and each block is Yang-Baxter
integrable. For particular loss/gain rates the model can be mapped to free
fermions. We determine the full quantum dynamics for an initial
product state in this case.
\end{abstract}
\noindent{\it Keywords\/}: open quantum systems, exact results, Lindblad equations, stochastic processes

\maketitle

\section{Introduction}
Whilst most standard tools for many-body quantum mechanics only apply
to closed systems, real systems are invariably influenced by their
environment. Under a Markovian approximation, an effective description
of this interaction can be given in terms of the Lindblad
equation~\cite{gorini1976completely,lindblad1976generators,breuer2002theory}
for the evolution of the density matrix. The main approaches that have
been used in the literature to study Lindblad equations for many body
systems are either perturbative
\cite{li2014perturbative,sieberer2016keldysh} or numerical
\cite{daley2014quantum,bonnes2014superoperators,Cui2015variational,Werner2016positive,weimer2021simulation}. 
Given that solvable models have provided deep insights
into the non-equilibrium dynamics of closed many body systems
\cite{calabrese2016introduction,essler2016quench,calabrese2016quantum,GHDreview1}
it is natural to ask if there are any exact results that can be obtained for
many particle Lindblad equations.

The first step in this direction was
the realisation that certain Lindblad equations can be cast in the
form of imaginary-time Schr\"odinger equations with non-Hermitian
``Hamiltonians'' that are quadratic in fermionic or bosonic field
operators~\cite{prosen2008third}, which then can be analyzed by
standard methods for free theories to extract physical properties
\cite{eisert2010noise,prosen2010spectral,clark2010exact,Horstmann2013noise,keck2017dissipation,vernier2020mixing,Somnath2020growth,alba2021spreading}. A
characteristic feature of these models is the fundamental boson or
fermion operators fulfil linear equations of motion
and concomitantly so do the Green's functions of interest. Another
step towards obtaining exact solutions of many particle Lindblad
equations was the discovery that there exist classes of models
in which some or all local correlation functions satisfy closed
hierarchies of equations of
motion~\cite{eisler2011crossover,vzunkovivc2014closed,caspar2016dissipative,Caspar2016dynamics,Hebenstreit2017Solvable,Foss-Feig2017solvable,klich2019closed}. This
permits one to obtain some exact results on the dynamics although full
solutions typically remain out of reach. Another class of solvable
Lindblad equations are ``triangular'' models which add particle loss
and dephasing terms to otherwise number conserving integrable models
\cite{Torres2014withoutgain,nakagawa2021exact,buca2020dissipative}.
Recently a new direction for 
constructing solvable many particle Lindblad equations was identified
through the discovery of Lindblad equations that can be related to
\emph{interacting} Yang-Baxter integrable
models~\cite{Medvedyeva2016,Rowlands2018noisy,Naoyuki2019dissipative,Naoyuki2019dissipativespin,Ziolkowska2020,nakagawa2021exact,buca2020dissipative,Ribeiro2019integrable,lerma2020trigonometric,yuan_solving_2021,deLeeuw21}. The
approach of Refs \cite{Medvedyeva2016,Ziolkowska2020} is based on a
superoperator representation of the Lindblad equations, which gives
rise to solvable ``two-leg ladder'' quantum spin chain
models. Importantly the equations of motion for correlation functions
do not generally close in these models but form an infinite
hierarchy of coupled nonlinear equations. More recently a method for
constructing Yang-Baxter integrable Lindblad systems was developed
\cite{deLeeuw21}. 

A related but different route
of constructing Yang-Baxter integrable Lindblad equations was
discovered in Ref.~\cite{PE20}. It is based on a ``fragmentation'' of
the space of operators into an exponential (in system size) number of
subspaces that are left invariant under the dissipative
evolution. Importantly, this mechanism applies to the quantum version
of the simple asymmetric exclusion process (ASEP)
\cite{SPITZER1970246,Liggett70,DERRIDA199865,Schutz00}. The
corresponding Lindblad equation can be obtained
\cite{Jin2020Stochastic} as the averaged dynamics of a stochastic
quantum model of particles hopping with random amplitudes first
introduced in its symmetric form in Ref.~\cite{Bauer2017Stochastic}
and further analyzed in
Refs.~\cite{Bauer2019Equilibrium,Bernard2019Open,bernard_solution_2020,frassek2020duality}. In this case, it was shown in \cite{PE20} that the Lindbladian restricted
to each invariant subspace can be mapped onto an XXZ Heisenberg
Hamiltonian with integrable boundary conditions. In particular, in the
subspace of diagonal density matrices the model reduces to the
classical ASEP, which is exactly solvable \cite{Gwa1992six,Gwa1992bethe} and
for which many exact results have been derived using integrability methods
\cite{Kim1995Bethe,golinelli2004bethe,de_Gier2005Bethe,de2006exact,de2008slowest,mallick2011some,crampe2011matrix}. While
\cite{PE20} established the integrability of the quantum ASEP in each
fragmented sector, the full solution of the dissipative quantum
dynamics remains an open problem except in the special case of the
quantum symmetric simple exclusion process. In order to show how the
operator-space fragmentation can be exploited in practice to obtain a
full solution of the dissipative dynamics we here consider a
generalization of the quantum ASEP. As we will show, the Lindbladian
of this model exhibits operator-space fragmentation and in each sector
can be mapped onto a Lindbladian that is quadratic in fermions. The
resulting dynamics can then be solved exactly.

The rest of this paper is organised as follows: in Section
\ref{sec:ASEP} we introduce the model of interest, which can be
viewed as an ASEP with additional loss/gain terms, and the model exhibits operator-space fragmentation, with each subspace labelled by a sequence of
``defects''. 
In Section \ref{sec:LossGain} we analyse the Lindbladian's projection on to each of these subspaces. We then focus on a particular line in parameter
space, on which the Lindbladian in each sector can be mapped onto a
bilinear form in auxiliary fermions. We show that the subspace of
diagonal density matrices is invariant under time evolution and
reduces to a classical stochastic process similar to ones that have
been previously studied in the literature
\cite{Santos96,Bares99,Bares2001,Crampe2016}. We employ Jordan-Wigner
and Bogoliubov transforms to solve the dynamics in this sector and
show that it has an infinite temperature steady state. In Section
\ref{sec:Defects} we consider the defect problem and outline how to
efficiently find the spectrum of the Lindbladian. 
In section \ref{sec:FM} we consider evolution out of an initial
product state and compute the transverse spin-spin correlation function.
Lastly, we relegate some technical calculations necessary for
the conclusions in the main text to two appendices. 
\section{Lindblad equation}
\label{sec:ASEP}
For a system interacting with its environment, the Lindblad equation for the time evolution of the reduced density
operator of the system $\rho$ is given by
\begin{eqnarray}
    \frac{\rmd \rho}{\rmd t} = -i[H,\rho]+\sum_a J_a\left(L_a\rho L_a^\dag - \frac{1}{2}\{L_a^\dag L_a,\rho\}\right),
    \label{Eq:Lindblad}
\end{eqnarray}
where the jump operators $L_a$ describe the interactions of the system
with the environment, $J_a$ are the corresponding rates and
$\{\cdot,\cdot\}$ denotes an anti-commutator. The Lindblad equation
\fr{Eq:Lindblad} describes the time evolution of the system degrees of
freedom after averaging over Markovian bath degrees of freedom
\cite{breuer2002theory}. In order to study the fluctuations of system
degrees of freedom that are induced by coupling to the bath --  a
question that has been extensively studied for classical systems (see
e.g. \cite{derrida1993exact,bertini2002macroscopic,bodineau04current,bertini2005current,Derrida2007Non-equilibrium,deGierEssler11,lazarescu2011exact}) -- it is
necessary to go beyond this description, see
e.g. \cite{Bauer2017Stochastic,Jin2020Stochastic,bernard_solution_2020},
but these fluctuations can still be described in terms of a quantum
master equation of Lindblad form \cite{bernard2020dynamics}. In
contrast, quantum measurement noise
\cite{gritsev2006full,kitagawa2011dynamics,eisler2013full,groha2018full,collura2019relaxation,collura2020how,javier2020relaxation}
is captured by the description \fr{Eq:Lindblad}. Since
\fr{Eq:Lindblad} is manifestly linear in $\rho$ it can be recast in
terms of a Lindblad superoperator that generates time evolution in the
same way the Hamiltonian does in closed quantum systems, with the
major difference being the time evolution need no longer be
unitary. That is, there exists a (super)operator ${\cal L}$ acting on
the vector space of linear operators on the Hilbert space such that
\begin{equation}
    \frac{\rm d}{{\rm d}t}|\rho\rrangle = {\cal L}|\rho\rrangle\ .
\end{equation}
Here we have written $|\rho\rrangle$ to stress that we are considering
$\rho$ as a vector in a larger vector space whose dimension is the
square of that for the original Hilbert space.
In this work we consider an open spin $1/2$ chain with periodic
boundary conditions and no coherent dynamics ($H=0$) described by four
jump operators \cite{EZ2020}
\begin{eqnarray}
L^{(1)}_j=\sigma^+_j\sigma^-_{j+1}\ ,\quad 
L^{(2)}_j=\sigma^-_j\sigma^+_{j+1}, \nn L^{(3)}_j=\sigma^+_j\sigma^+_{j+1}\ ,\quad
L^{(4)}_j=\sigma^-_j\sigma^-_{j+1}.
\label{jumpops}
\end{eqnarray}
In terms of Jordan-Wigner fermions the first two of these correspond
to hopping left and right, whilst the latter two represent pair
creation and annihilation on neighbouring sites respectively. In 
general the rates of these may all be different and one obtains a four
parameter family of models \cite{EZ2020}. The case $J_3=J_4=0$ reduces
to the quantum ASEP \cite{Bauer2017Stochastic,Jin2020Stochastic,PE20}.
In contrast to the latter case the additional jump
operators describe processes that violate spin rotational invariance
around the $z$-direction (or equivalently particle number conservation
at the level of Jordan-Wigner fermions) so that the magnetization
is no longer conserved. As we will see this leads to interesting new
effects compared to the ASEP. The Lindblad equation \fr{Eq:Lindblad} with jump
operators \fr{jumpops} can be obtained by coupling our quantum spins
across each bond of our chain to an environment modelled by
appropriate quantum Brownian motions as in \cite{Jin2020Stochastic}
and then averaging over the bath degrees of freedom. Our choice of
model is not motivated by any particular experimental setup, but aims
to address a problem in mathematical physics, namely to obtain a
many-particle Lindblad equation exhibiting operator-space
fragmentation that can be solved exactly in practice. Having said
this, in a particular parameter regime and for diagonal initial
density matrices our model reduces to a classical master equation that
has been argued to describe the kinetics of excitons in certain polymers
\cite{Santos96} and it would be interesting to investigate whether
quantum effects could be relevant to this system.
In order to recast the Lindblad equation \fr{Eq:Lindblad} with jump
operators \fr{jumpops} in the superoperator formalism we note that the
density operator is expressed as
\begin{equation}
    \rho=\sum_{\alpha,\beta} \rho^{\alpha}_{\; \beta}|\alpha\rangle \langle \beta | \mapsto |\rho \rrangle = \sum_{\alpha \beta} \rho^{\alpha \beta}|\alpha \rangle \otimes |\beta \rangle \label{Eq:Super-Conventions}.
\end{equation}
Then right multiplication by an operator $L$ must turn into left
multiplication by some superoperator $L_R$ such that
\begin{eqnarray}
    L_R |\rho\rrangle  = \sum_{\alpha \beta} \left(\rho^{\alpha
      \gamma}L^{\;\beta}_\gamma\right) |\alpha\rangle \otimes |\beta
    \rangle\ ,
\end{eqnarray}
which implies that $L_R=\id \otimes \left(L^{\;
  \beta}_{\gamma}|\beta\rangle \langle \gamma |\right)$. Note that
this has indices swapped compared to \ref{Eq:Super-Conventions},
indicating that the right multiplication action is implemented via the
transpose of the original operator, along with acting on bras instead
of kets. Left multiplication is simply implemented via the operator
acting on kets. This can be summarised as $L_L=L\otimes \id$ and
$L_R=\id \otimes L^T$.
In order to obtain an explicit expression for the Lindblad superoperator
${\cal L}$ we pick the following basis of the local Hilbert space of
operators acting on site $j$
\begin{equation}
    |1\rrangle_j = |\! \uparrow\rangle_j\ {}_j\langle \uparrow \!|, \quad 
    |2\rrangle_j = |\! \uparrow\rangle_j\ {}_j\langle  \downarrow\!|, \quad
    |3\rrangle_j = |\! \downarrow\rangle_j\ {}_j\langle \uparrow\! |, \quad
    |4\rrangle_j = |\! \downarrow\rangle_j\ {}_j\langle \downarrow\! |.   
\end{equation}
A basis of local superoperators acting on these states in then given
by
\be
E_j^{ab}=|a\rrangle_j\ {}_j\llangle b|\ .
\ee
For convenience, we split the Lindbladian up as
\be
{\cal L}={\cal  L}^{\rm Diag}+{\cal L}^{\rm Defect}\ ,
\label{Ltotal}
\ee
where ${\cal L}^{\rm Diag}$ leaves invariant the subspace of diagonal density
matrices

\be
|\rho\rrangle_{\rm diag}=\sum_{\boldsymbol{\sigma}}
\rho^{\boldsymbol{\sigma}\boldsymbol{\sigma}}|\boldsymbol{\sigma}\rangle\otimes|\boldsymbol{\sigma}\rangle\ ,\quad
|\boldsymbol{\sigma}\rangle=\otimes_{j=1}^L|\sigma_j\rangle_j\ ,
\sigma_j\in\{\uparrow,\downarrow\}.
\label{rhodiag}
\ee
These diagonal density matrices correspond to classical probability distributions. We have 
\begin{align}
{\cal L}^{\rm Diag}&= \sum_j J_1 E_j^{14}E_{j+1}^{41} + J_2E_j^{41}E_{j+1}^{14} + J_3 E_j^{14}E_{j+1}^{14} + J_4 E_j^{41}E_{j+1}^{41} \nn 
&-\sum_jJ_1E_j^{44}E_{j+1}^{11}+J_2E_j^{11}E_{j+1}^{44}
+J_3E_j^{44}E_{j+1}^{44}+J_4E_j^{11}E_{j+1}^{11}\ ,
\label{Eq:L_Diag_Es}\\
{\cal L}^{\rm Defect}&=
-\frac{1}{2}\sum_j(E_j^{22}+E_j^{33})([J_1+J_4]E_{j+1}^{11}+[J_2+J_3]E_{j+1}^{44})\nn
& -\frac{1}{2}\sum_j(E_{j+1}^{22}+E_{j+1}^{33})([J_2+J_4]E_{j}^{11}+[J_1+J_3]E_{j}^{44})\nn
&  -\frac{J_1+J_2+J_3+J_4}{2}(E_j^{22}E_{j+1}^{33}+E_j^{33}E_{j+1}^{22}).
\label{Eq:L_Defect_Es}   
\end{align}
If we initialize the system in a purely diagonal density
matrix the Lindblad equation \fr{Eq:Lindblad} reduces to a classical
master equation with transition matrix ${\cal L}^{\rm Diag}$. This
describes generalizations of the asymmetric simple exclusion process
\cite{SPITZER1970246,Liggett70,DERRIDA199865,Schutz00} similar to the
diffusion-annihilation models studied in
\cite{Santos96,Bares99,Bares2001,Crampe2016}. If we set $J_3=J_4=0$ we
recover the ASEP with periodic boundary conditions.
\subsection{Operator-space fragmentation}
The origin of operator-space fragmentation in the model \fr{Ltotal},
\fr{Eq:L_Diag_Es}, \fr{Eq:L_Defect_Es} is the presence of strictly
local conservation laws 
\begin{equation}
    [{\cal L},E_j^{22}]=0=[{\cal L},E_j^{33}].
\label{commute}
\end{equation}
These conservation laws imply that particles of species $2,3$ are left
invariant by the dynamics and we therefore refer to these as ``defects''. The
Hilbert space of operators thus breaks up into exponentially many invariant subspaces with
fixed occupancies of defects. This is somewhat reminiscent of the Hilbert space
fragmentation found in certain fractonic
circuits \cite{Sala2020Ergodicity,Khemani2020Localization,Moudgalya2019thermalization}.
The fragmentation of operator-space does not rely on the fact our
model is one dimensional. Indeed, operator-space fragmentation occurs
if we consider a square lattice and jump operators defined on all
nearest neighbour bonds
\be
     L_j^{(u)} = \sigma^+_{(i,j)}\sigma^-_{(i,j+1)} \ , \quad
     L_j^{(d)} = \sigma^-_{(i,j)}\sigma^+_{(i,j+1)}\ .
\ee
In this case the $2L^2$ operators $E_{(i,j)}^{22},E^{33}_{(i,j)}$ are
then strictly conserved. By focusing on one dimensional models however
we allow for the possibility that the Lindbladian's action on each
subspace can be mapped to an integrable model. However, the ocurrence of
fragmentation will have implications for the dynamics in higher
dimensions as well.

This operator-space fragmentation then
allows observables to be computed by analyzing each sector
separately. In the case of the ASEP ($J_3=J_4=0$), the key result is
that restricted to each defect-subspace the Lindbladian can be mapped
to a collection of disjoint finite XXZ chains with diagonal boundary
fields and is thus integrable on every subspace. Integrability
techniques can be similarly applied to \fr{Ltotal}, \fr{Eq:L_Diag_Es},
\fr{Eq:L_Defect_Es} \cite{Crampe2016,EZ2020} but we do not pursue this line of
enquiry here and instead impose a particular constraint on the rates
$J_1,\dots,J_4$ which will allow us to employ mappings to free fermion
systems (see below).

It should be stressed that for a particular observable, it may not be
necessary to deal with very many invariant subspaces. This is
illustrated by the transverse spin-spin correlation function 

\begin{equation}
    S^{+-}_{0,\ell+1}=\Tr \left[\sigma_0^+ \sigma_{\ell+1}^- \rho(t) \right].
\end{equation}
This depends only on the subspace with a type $3$ defect at site $0$ (equivalently site $L$) and a type $2$ defect at site $\ell+1$. To see this, note that in the superoperator formalism traces are replaced by inner products with the state
\begin{equation}
\llangle \idc| =\otimes_{j=1}^L \left[{}_j\llangle 1| + {}_j\llangle 4|\right].
\end{equation}
An immediate consequence of the fact that the time evolution operator
$e^{{\cal L}t}$ preserves traces is that $\llangle \idc |$ is a left
eigenvector of the time evolution operator with eigenvalue $1$. If
there is a unique steady state of the system then it is also the only
left eigenvector with this property, a fact that we will use
later. The spin operators act by left multiplication so in the
  superoperator formalism they are mapped to 
\begin{eqnarray}
\sigma_0^+ \mapsto \sigma_0^+ \otimes \id_0 = (E_0^{13}+E_{0}^{24})\ , \nn
\sigma_{\ell+1}^- \mapsto \sigma_{\ell+1}^- \otimes \id_{\ell+1} = (E_{\ell+1}^{42}+E_{\ell+1}^{31})\ .
\label{Spm}
\end{eqnarray}
Since $\llangle \idc|$ only contains states $1,4$ the only terms that survive in the trace are then
\begin{equation}
    S^{+-}_{0,\ell+1}= \Big[{}_0\llangle 3|\otimes\llangle \id_{[1,\ell]}|\otimes
    {}_{\ell+1}\llangle 2|\otimes\llangle \id_{[\ell+2,L-1]}|\Big]| \rho(t)\rrangle \ ,
\label{Spm2}
\end{equation}
where we have introduced
\be
\llangle\id_{[a,b]}|=\otimes_{j=a}^b\left[{}_j\llangle 1| + {}_j\llangle 4|\right].
\ee
Eqn \fr{Spm} shows that the correlation function depends only on the
projection of $|\rho(t)\rangle$ on the single subspace described
above. This means the correlation function can be written in terms of
propagators defined on open chain segments: 
\begin{equation}
    G_{[a,b]} = \llangle \id_{[a,b]}|e^{{\cal L}_{[a,b]}t}|\rho_{[a,b]}\rrangle.
\end{equation}
In the ASEP case these propagators involve computing the overlap of a time evolved state in the finite length XXZ model (with diagonal boundary fields) with the state $\langle \idc|$. The rest of this paper will consider a different subspace of the full four-parameter model which reduces to free fermions, thus allowing the calculation of $G_{[a,b]}$ for some initial states, although its calculation for general states is still difficult .

\section{Free fermions}
\label{sec:LossGain}

\subsection{``Classical'' sector}
As we noted earlier, the subspace of diagonal density matrices
\fr{rhodiag} is invariant under the dynamics. The ``classical'' part
${\cal L}^{\rm Diag}$ of the Lindbladian acts on this $2^L$
dimensional subspace of diagonal density matrices and can be expressed
in terms of Pauli matrices $\tau_j$ defined by
\be
    \tau^z_j = E_j^{11}-E_j^{44}\ , \quad
    \tau^+_j = E_j^{14}\ .
\label{classicalspins}
\ee
We find
\begin{align}
{\cal L}^{\rm Diag}&= \sum_j J_1 \tau^+_j \tau^-_{j+1} + J_2\tau^-_j \tau^+_{j+1} + J_3 \tau^+_j \tau^+_{j+1} + J_4 \tau^-_j \tau^-_{j+1}   \nn 
&    -\frac{1}{4}\sum_j (J_1-J_2-J_3+J_4)\id_j \tau^z_{j+1} +
(-J_1+J_2-J_3+J_4)\tau^z_{j} \id_{j+1} \nn
&-\frac{1}{4}\sum_j (J_1 + J_2 - J_3 - J_4)\tau_j^z\tau_{j+1}^z + (J_1+J_2+J_3+J_4) \id_j\id_{j+1}.
    \label{Eq:L_Diag}
\end{align}
We now observe that under the constraint
\be
J_1+J_2 = J_3+J_4,
\label{constraint}
\ee
the model \fr{Eq:L_Diag} can be mapped to a free fermionic theory by
means of a Jordan-Wigner transformation. In the periodic case we use
that $\sum_j^L \id_j\tau^z_{j+1}-\tau^z_j\id_{j+1}=0$ to obtain 
\bea
\fl {\cal L}^{\rm Diag}=\sum_{j=1}^{L-1}\Big\{J_1
c^\dagger_{j+1}c_j + J_2 c^\dagger_j c_{j+1}-J_3 c_jc_{j+1} -J_4 c^\dagger_{j+1}c^\dagger_j
+(J_4-J_3)c^\dagger_jc_j \big\} \nn + (-1)^{\hat{N}}\left(J_1c_{1}^\dag c_L + J_2 c_L^\dag c_1 - J_3 c_1^\dag c_L^\dag - J_4 c_L c_1\right)-J_4L,
\eea
where $\hat{N}$ is the total fermion number operator. Since each term
in the Lindbladian preserves fermion parity, the operator
$(-1)^{\hat{N}}$ is conserved and we can work in definite parity
sectors where it equals $+1$ (periodic, or Ramond, boundary
conditions) or $-1$ (anti-periodic, or Neveu-Schwarz, boundary
conditions). It will furthermore be convenient in the following to
define
\bea
    2J_+ &= J_1+J_2=J_3+J_4\ , \nn
    2J_- &= J_1-J_2\ , \nn
    2\mu &= J_4-J_3\ ,
\label{Jpm}
\eea
in terms of which the Lindbladian can be written (defining
$c_{L+1}=(-1)^{\hat{N}}c_1$) as 
\begin{align}
{\cal L}^{\rm Diag}=(J_++\mu)L + \sum_{j=1}^{L}&\Big\{(J_++J_-)
c^\dagger_j c_{j+1}+(J_+-J_-) c^\dagger_{j+1}c_j\nn
&-\big[ (J_++\mu) c_j c_{j+1} + (J_+-\mu) c^\dagger_{j+1}c^\dagger_j\big]
+2\mu c^\dagger_jc_j \Big\}.
\end{align}
We largely focus on the special case $\mu=0$ in the following but do discuss
the steady state in the imbalanced case in Section
\ref{Sec:Imbalanced}. Crucially the constraints \fr{Jpm}
enforce that $J_3+J_4\neq 0$, which takes us away from the ASEP limit
$J_3=J_4=0$. Hence the exact solutions presented below cannot be related
to known results for the ASEP.

\subsection{Two defect sector}
We now consider the case where there are two defects that without loss
of generality can be taken to be located at positions $\ell+1$ and
$L$. Inspection of \fr{Ltotal}, \fr{Eq:L_Diag_Es}, \fr{Eq:L_Defect_Es}
shows that on the corresponding subspace the Lindbladian takes the form
\be
{\cal L}=\begin{cases}
{\cal L}_{[1,\ell]}+{\cal L}_{[\ell+2,L-1]} & \text{if }0<\ell<L-1\ ,\\
{\cal L}_{[2,L-1]}+c
& \text{if }\ell=0\ ,\\
{\cal L}_{[1,L-2]}+c& \text{if }\ell=L-2\ ,
\end{cases}
\ee
where
\begin{align}
{\cal L}_{[1,\ell]}&= \sum_{j=1}^{\ell-1}
  J_1 E_j^{14}E_{j+1}^{41} + J_2E_j^{41}E_{j+1}^{14} + J_3 E_j^{14}E_{j+1}^{14} + J_4 E_j^{41}E_{j+1}^{41} \nn 
&-\sum_{j=1}^{\ell-1}J_1E_j^{44}E_{j+1}^{11}+J_2E_j^{11}E_{j+1}^{44}
+J_3E_j^{44}E_{j+1}^{44}+J_4E_j^{11}E_{j+1}^{11}\nn
&-\frac{1}{2}\Big([J_1+J_4]E^{11}_{1}+[J_2+J_3]E^{44}_{1}+[J_2+J_4]E^{11}_\ell+[J_1+J_3]E^{44}_{\ell}\Big),
\end{align}
and the constant $c$ is given by $c=-(J_1+J_2+J_3+J_4)/2=-2J_+$ if one of the defects is of type $2$ and one of type $3$  and zero if the two defects are of the same type. Imposing the constraint \fr{constraint} and carrying out a
Jordan-Wigner transformation to spinless fermions we arrive at a free
fermion chain with open boundary conditions
\begin{align}
{\cal L}_{[1,\ell]} &= -J_+(\ell+1)-\mu\sum_{j=1}^\ell (2c^\dagger_jc_j-1)\nn
&+ \sum_{j=1}^{\ell-1}\Big\{J_1 c^\dagger_{j+1}  c_{j}+J_2 
c^\dagger_{j}c_{j+1} -J_3 c_{j}c_{j+1}
-J_4 c^\dag_{j+1}c^\dag_{j} \big)  \Big\}. 
\label{Eq:Liouv_2defect}
\end{align}
\subsection{\sfix{$q$} defect sector}
The Lindbladian for the entire chain restricted to the invariant subspace with $q$ defects at locations $\ell_1,\dots \ell_q$ is simply a sum of Lindbladians for the $q$ disjoint finite chains obtained by removing the sites $\ell_j$ from the original chain
\be
    {\cal L} = \sum_{j=0}^q {\cal L}_{[\ell_{j}+1,\ell_{j+1}-1]}
\ee
Here $\ell_0=\ell_q$ so that for instance in the $1$ defect sector the
corresponding Lindbladian ${\cal L}_{[\ell+1,\ell-1]}$ corresponds to the
original ring with a single site removed. If the defects
$\ell_j,\ell_{j+1}$ are not immediate neighbours then these are
exactly as given in Eq \fr{Eq:Liouv_2defect}. If there are two
neighbouring defects then the only term in the full Lindbladian that
acts on them is 
\be
    {\cal L}_{[\ell,\ell+1]} = -2J_+ (E^{22}_{\ell}E^{33}_{\ell+1}+E^{33}_{\ell}E^{22}_{\ell+1})
\ee
which contributes $c=-2J_+$ if the neighbouring defects are different species and $0$ if they are the same.

\section{Dynamics in the classical subspace}
As a first step to understanding this model we solve it exactly in the diagonal subspace. We focus initially on the balanced ($\mu=0$) case. Our system has periodic boundary conditions in terms of the original spins and (anti)-periodic boundary conditions for Jordan-Wigner fermions in sectors of (even) odd fermion parity. We therefore go to Fourier space
\be
c(k_n)=\frac{1}{\sqrt{L}}\sum_je^{ik_nj} c_j\ ,\quad k_n=\frac{2\pi (n+\delta)}{L}.
\ee
where $\delta=0,1/2$ for states with odd or even fermion parity respectively. We then carry out a Bogoliubov transformation to diagonalize the Lindbladian
\begin{eqnarray}
c^\dagger(k)&=\cos(k/2) b_{-k}-i\sin(k/2) b^\dagger_k\ ,\nn
c(k)&=i\sin(k/2) b_k+\cos(k/2) b^\dagger_{-k}\ .
\end{eqnarray}
Despite the fact that ${\cal L}$ is non-Hermitian, this transformation is still unitary. We have
\be
{\cal L}^{\rm Diag}=\sum_{k}\epsilon(k) b^\dagger_k b_k\ ,
\label{Lioudiag}
\ee
where the non-Hermitian nature of the Lindbladian presents through the complex eigenvalues
\begin{equation}
\epsilon(k)=-2J_++2iJ_-\sin k\ .
\end{equation}
The time-evolved operators are
\be
b_k(t)=e^{-{\cal L}^{\rm Diag}t}b_ke^{{\cal L}^{\rm Diag}t}=e^{\epsilon(k)t}b_k\ .
\ee
We can now immediately conclude that the stationary state is unique
and given simply by the Bogoliubov vacuum 
\be
b_k|0\rrangle=0.
\ee
This implies that $\llangle 0|{\cal L}=0$ and exploiting uniqueness we
therefore have
\be
\llangle \idc |=\llangle 0|\ .
\ee
This is turn shows that the stationary state $|0\rrangle$ is the
completely mixed (infinite temperature) state, which we now demonstrate in
more detail.  

An important question is what operators of the
  original spin-chain problem can have finite expectation values
  within the defect-free subspace. To answer this we project the
  original Pauli matrices 
$\sigma_j$ on to the diagonal subspace and write the result in terms of
the $\tau_j$ operators. Defining projection operators by
\be
P_j=E_j^{11}+E_j^{44},
\ee
we have
\begin{eqnarray}
P_j\left[\sigma^z_j \otimes \id_j\right]P_j
&= P_j\left[E^{11}-E^{44}+E^{22}-E^{33}\right]P_j
=\tau^z_j=1-2n_j\ ,\nn
P_j\left[\sigma^\alpha_j \otimes \id_j\right]P_j&=0\ ,\quad \alpha=x,y.
\end{eqnarray}
This shows that the only physical operators with non-zero expectation
in the stationary state are 
\be
{\cal O}_{j_1,\dots,j_n}=n_{j_1}\dots n_{j_n}\ . 
\ee
The expectation value of ${\cal O}_{j_1,\dots,j_n}$ can be obtained
using Wick's theorem with the help of the elementary two-point functions
\begin{eqnarray}
\llangle 0|c^\dagger_jc_{j+\ell}|0\rrangle&=\frac{\delta_{\ell,0}}{2}+\frac{\delta_{\ell,1}+\delta_{\ell,-1}}{4},\nn
\llangle 0|c_jc_{j+\ell}|0\rrangle&=\frac{\delta_{\ell,-1}-\delta_{\ell,1}}{4},\nn
\llangle 0|c^\dagger_jc^\dagger_{j+\ell}|0\rrangle&=\frac{\delta_{\ell,1}-\delta_{\ell,-1}}{4}.
\end{eqnarray}
Here we have replaced the state $\llangle \idc |$ used to compute traces with the left Bogoliubov vacuum $\llangle 0|$ following the discussion above. As a result, we find that all such expectations factorise
\be
\llangle 0|{\cal O}_{j_1,\dots,j_n}|0\rrangle=\llangle 0|n_{j_1}|0\rrangle
\dots\llangle 0| n_{j_n}|0\rrangle=\frac{1}{2^n}\ . 
\ee
We now make use of the fact that a density operator is fully
determined by the expectation values of a complete set of operators to
conclude that in terms of the original problem the stationary state is
the infinite temperature state  
\be
\rho_{\rm
  stat}= \frac{1}{2^L} \sum_{\sigma_1,\dots,\sigma_L}|\sigma_1,\dots,\sigma_L\rangle\langle
\sigma_1,\dots,\sigma_L|.
\ee
\subsection{Imbalanced loss and gain}
\label{Sec:Imbalanced}
We now briefly discuss the nature of the steady state with imbalanced loss and gain. When $\mu\neq 0$ the Lindbladian in the defect-free sector is given by
\be
    {\cal L}=-(J_++\mu)L+\sum_{j=1}^L\left\{J_1c_{j+1}^\dag c_j + J_2 c_j^\dag c_{j+1} -J_3 c_j c_{j+1}-J_4 c_{j+1}^\dag c_j^\dag +2\mu c_j^\dag c_j  \right\} \ ,
\ee
where the appropriate Ramond or Neveu-Schwarz boundary conditions are
assumed. We make use of the translational symmetry in the defect-free
problem to Fourier transform this to give 
\be
    {\cal L}={\rm const} + \sum_{k>0} \vec{c}_k^\dag A_k(\mu) \vec{c}_k \ , 
\ee
where $\vec{c}_k=\begin{pmatrix}c_k & c^\dag_{-k}\end{pmatrix}^T$ the matrix $A_k(\mu)$ is $2\times 2$ and non-Hermitian
\be
    A_k(\mu) = 2J_+\begin{pmatrix}i\Delta \sin k+\cos k + \nu & -i(1+\nu)\sin k \cr +i(1-\nu)\sin k & i\Delta \sin k - \cos k - \nu \end{pmatrix}\ .
\ee
Here we have introduced the dimensionless parameters
\be
    \nu = \frac{\mu}{J_+} = \frac{J_4-J_3}{J_4+J_3} \quad \Delta = \frac{J_-}{J_+}=\frac{J_1-J_2}{J_1+J_2}\ ,
\label{dimlessparameters}
\ee
The parameter $\nu$ satisfies $-1\leq \nu \leq 1$ where the extreme case of $\nu=-1$ corresponds to only particle loss and $\nu=1$ to only gain. In terms of $\nu$ the eigenvalues of $A_k(\mu)$ are
\be
    \epsilon_k^{\pm}(\mu) = 2J_+(i\Delta \sin k \pm (1+\nu \cos k) ) \ .
\ee
For $\nu \neq \pm 1$ these are always distinct. Degerate eigenvalues only occur for $\nu=1, k=\pi$ and $\nu=-1,k=0$ which both yield $A_k(\mu)=0$. $A_k(\mu)$ is thus always diagonalisable, however it is not unitarily diagonalisable if $\mu\neq 0 $. In this case we cannot perform a canonical transformation as for the balanced case.

We can however perform an almost canonical transformation by defining the matrix
\be
    S_k = \frac{1}{\sqrt{1+\nu \cos k}}\begin{pmatrix}   \sin \frac{k}{2} & -i(1+\nu)\cos\frac{k}{2} \cr -i\cos \frac{k}{2} & (1-\nu)\sin\frac{k}{2} \end{pmatrix},
\ee
chosen such that $S^{-1}AS$ is diagonal and ${\det}(S)=1$. We then define
\bea
\begin{pmatrix}
b'_{+,k} & b_{-,k} 
\end{pmatrix} = \begin{pmatrix}
        c_k^\dag & c_{-k}
    \end{pmatrix}S_k\ , \\
    \begin{pmatrix}
        b_{+,k} \cr b'_{-,k}
    \end{pmatrix} = 
    S^{-1}_k
    \begin{pmatrix}
        c_k \cr c_{-k}^\dag
    \end{pmatrix} \ .
\eea
These are almost canonical fermions in that they satisfy the relations
\bea
    \{b_{\sigma,k},b_{\tau,q}\} =& 0
    =\{b^{\prime}_{\sigma,k},b^{\prime}_{\tau,q}\}\ , \nn
    \{b^{\prime }_{\sigma,k},b_{\tau,q}\} =& \delta_{\sigma,\tau}\delta_{k,q} \ .
\eea
We note that $b_{+,-k}=-b_{-,k}$ due to the choice of normalisation in
the definition of $S_k$, which allows us to consistently define
\bea\fl
b_k =\theta(k)b_{+,k}+\theta(-k) b_{-,k}=
 \frac{{\rm sgn}\ k}{\sqrt{1+\nu\cos k}}
\Big( (1-\nu)\sin\frac{k}{2} c_k +i(1+\nu)\cos\frac{k}{2} c_{-k}^\dag
\Big)\ , \nn
\fl
b'_k = \theta(k)b^\prime_{+,k}+\theta(-k) b^\prime_{-,k}=
\frac{{\rm sgn}\ k}{\sqrt{1+\nu\cos k}}\Big( -i\cos \frac{k}{2}c_{-k}
+\sin\frac{k}{2} c_{k}^\dag \Big) \ .
\label{bbprime}
\eea
This then allows us to write the Lindbladian in terms of the
almost canonical fermion operators as  
\be
    {\cal L} = {\rm const} + \sum_{k}\epsilon_k^- b'_k b_k\ ,
\ee
where we have used that $\epsilon_{k}^+=-\epsilon_{-k}^-$.
The constant can be seen to be $0$ by carefully keeping track of the
constants discarded throughout this argument. We can now define left
and right vacua by 
\be
    \forall k: \quad \llangle L | b'_{k} = 0 , \quad b_{k}|R\rrangle = 0.
\ee
Since $\llangle L|{\cal L}=0$ has only one solution, we conclude that
\be
\llangle L | =\llangle\idc|= {}_0\llangle  0 |  \prod_{k>0} \Big( 1+i\cot
\frac{k}{2}c_{-k} c_{k}\Big)\ ,
\label{leftvac}
\ee
where $|0\rrangle_0$ is the fermionic vacuum state
\be
    c_j|0\rrangle_0=0\ .
\ee
The expression {\color{red}for} $\llangle L|$ in terms of the original
fermions in \fr{leftvac} is easily verified by acting with $b'_k$ and using \fr{bbprime}.
The right eigenstate can be expressed as a squeezed state via
\bea
    |R\rrangle &=\frac{1}{\cal N}\prod_{k>0}\Big(1-
    i\frac{1+\nu}{1-\nu}\cot \frac{k}{2} c_k^\dag c_{-k}^\dag \Big) |0\rrangle_0 ,
\label{SSmu}
\eea
where ${\cal N}$ is chosen such that $\llangle
L|R\rrangle=1$. This can be verified by acting with
$b_k$ and using \fr{bbprime}. 

As before we may consider the expectation values of all operators in
the classical subspace - the operators $\sigma^x,\sigma^y$ project to
zero and all physical operators are given in terms of fermions by
products of densities 
\begin{equation}
    {\cal O}_{j_1\dots j_r} = n_{j_1}\dots n_{j_r}\ .
\end{equation}
The simplest such expectation value is
\be
\llangle n_j\rrangle_\infty=\lim_{t\to\infty}{\rm
  Tr}\big[n_j\ \rho(t)\big]=\llangle L|n_j|R\rrangle. 
\ee
As we
have seen above, for $\mu=0$ the steady state corresponds to a
completely mixed state and we have $\llangle
n_j\rrangle_{\infty}=1/2$. When $\mu\neq 0$ and assuming that the
steady state is uncorrelated we have the following relation expressing
the balance between particle gain and loss
\be 
    J_3\llangle n_j\rrangle_\infty^2 = J_4[1-\llangle n_j \rrangle_\infty]^2.
\ee
If this relation holds we may solve for the particle density
\be
    \llangle n_j \rrangle_\infty = \frac{1+\nu-\sqrt{1-\nu^2}}{2 \nu} \ ,
\label{sdens}
\ee
where we have used \fr{dimlessparameters}.
We now verify \fr{sdens} by direct calculation. Due to translation
invariance $\llangle n_j\rrangle_\infty$ is the same on each site and we
can instead calculate the average occupation of the $k$ modes 
\be
    \frac{1}{L}\sum_k \llangle n_k\rrangle = \frac{1}{L}\sum_k
    \frac{1+\nu}{1+\nu \cos k}\cos^2\frac{k}{2}\llangle L | b_{-k}
    b'_{-k} |R\rrangle \ .
\ee
In the thermodynamic limit this turns into an integral
\begin{align}
\llangle n_m \rrangle_\infty  
&=  \frac{(1+\nu)}{4\pi}\oint \frac{1+\cos k}{1+\nu \cos k} \ {\rm d}k\nn
&= \frac{1+\nu-\sqrt{1-\nu^2}}{2 \nu} \ .
\end{align}
This indeed agrees with \fr{sdens}. We note that the stationary state
\fr{SSmu} has a simple product form in terms of the spin states $|1\rangle$, $|4\rangle$ on
which the spin operators $\tau_j^\alpha$ act, \emph{cf.} \fr{classicalspins}.
\begin{align}
\llangle L|&={\displaystyle\otimes_{j=1}^L}\left[{}_j\llangle 1|+{}_j\llangle
  4|\right]\ ,\nn
|R\rrangle&=\frac{1}{(1+\alpha)^{L}}\otimes_{j=1}^L\left[|1\rrangle_j+\alpha|4\rrangle_j\right]\ ,
\end{align}
where
\be
\alpha=\sqrt{\frac{1+\nu}{1-\nu}}\ .
\ee

\subsection{Time dependence}
\begin{figure}[ht]
\caption{Relaxation of $n_j(t)$ for odd/even sites towards the steady state value from an initial N\'eel state. Calculated for $J_+=1.0, J_-=0.9$.}
\centering
\includegraphics[width=0.75\textwidth]{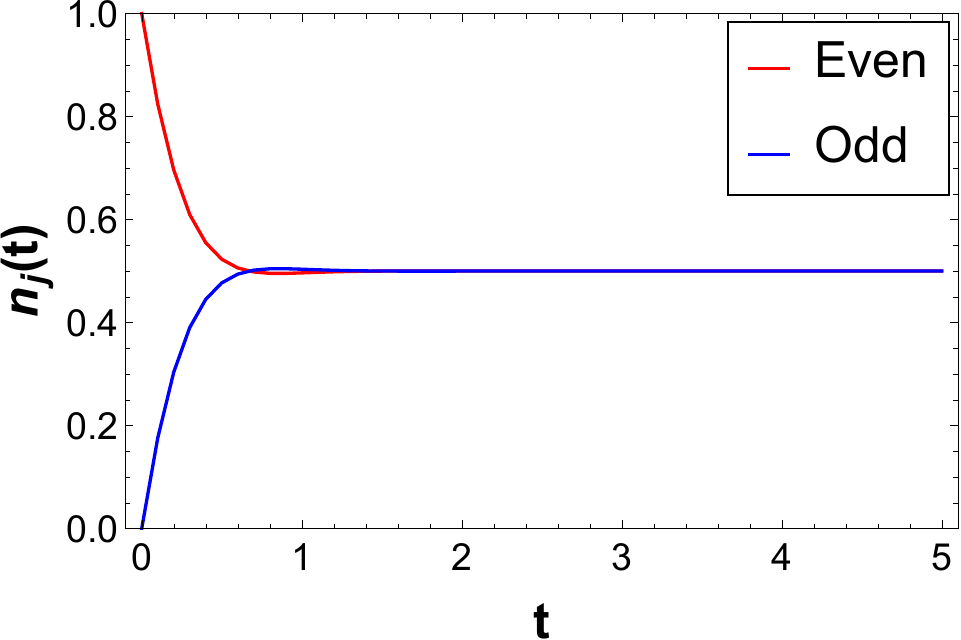}
\label{Fig:Neel}
\end{figure}

We return to considering only the balanced case of $\mu=0$ and now consider the time dependent problem. As we have seen above, on the diagonal subspace we have
\be
P_j\left[\sigma^z_j\otimes\id\right]P_j= 1-2c^\dagger_jc_j=
P_j\left[\id \otimes\tilde{\sigma}^z_j\right]P_j
\ee
This allows us to identify
\be
|\sigma_1,\dots,\sigma_L\rangle\langle \sigma_1,\dots,\sigma_L|=c^\dagger_{j_1}\dots
c^\dagger_{j_n}|0\rrangle_0\ ,
\label{statesid}
\ee
where $j_k$ are the positions of down spins ordered such that $j_1 < j_2
\dots <j_n$ and $|0\rangle_0$ is the fermionic vacuum, which is related to the Bogoliubov vacuum state by
\be
|0\rrangle_0=2^{-L}\prod_{k>0}\left[1+i\cot(k/2)b^\dagger_kb^\dagger_{-k}\right]|0\rrangle.
\ee
Using this and an initial density matrix in our subspace we can
calculate 
\be
\llangle \idc|n_{j_1}(t)\dots n_{j_n}(t)|\rho(0)\rrangle.
\ee
We can thus compute the expectations of any observables in this
subspace at arbitrary times using free-fermion techniques. As an
example we now compute $\llangle n_j(t)\rrangle $ for a
system initially in the classical N\'eel state
\begin{equation}
    \rho(0)=|\uparrow \downarrow \uparrow \downarrow \dots\rangle\langle\uparrow\downarrow\uparrow\downarrow\dots|.
\end{equation}
In terms of fermions this can be written as
\be
|\rho_{\text{N\'eel}}\rrangle=\prod_{j=1}^{N/2} c^\dagger_{2j}|0\rrangle_0.
\ee
In practice it will be more useful to work with the original fermion operators than the Bogoliubov ones. Solving their equations of motion gives
\begin{eqnarray}
c^\dagger(k,t)&= f(k,t)c^\dagger(k)+g(k,t)c(-k)\ ,\nn
c(-k,t)&= -g(k,t)c^\dagger(k)+h(k,t)c(-k)\ ,
\end{eqnarray}
where
\begin{eqnarray}
f(k,t)&=\cos^2(k/2)e^{\epsilon(-k)t}+\sin^2(k/2)e^{-\epsilon(k)t}\ ,\nn
g(k,t)&=\frac{i}{2}\sin(k)\left[e^{\epsilon(-k)t}-e^{-\epsilon(k)t}\right],\nn
h(k,t)&=\cos^2(k/2)e^{-\epsilon(k)t}+\sin^2(k/2)e^{\epsilon(-k)t}\ .
\end{eqnarray}
This then allows us to write
\begin{align}
n_j(t)&=\frac{1}{L}\sum_{p,q}e^{ij(p-q)}c^\dagger(p,t)c(q,t)\nn
&=\sum_{m}\left[\tilde{f}_{m-j}(t)c^\dagger_m+\tilde{g}_{m-j}(t)c_m\right]
\sum_n\left[\tilde{h}_{n-j}(t)c_n-\tilde{g}_{n-j}(t)c^\dagger_n\right],
\end{align}
where we have defined
\be
\tilde{f}_n(t)=\frac{1}{L}\sum_p e^{-ipn}f(p,t)\ .
\ee
To calculate $\langle\id_{\rm cl}|n_j(t)|\rho_{\text{N\'eel}}\rangle$ it is
helpful to split the double sum in $n_j(t)$'s Fourier series into 
\newcommand{\odd}{\ {\rm odd}}
\newcommand{\even}{\ {\rm even}}
\begin{equation}
     \sum_{m,n} = \sum_{n \even}\delta_{mn}+\sum_{n \odd}\delta_{mn} + \sum_{\substack{n \even \cr m \neq n}}+\sum_{\substack{n \odd \cr m\neq n}}.
 \end{equation}
A straightforward calculation then gives
 \begin{eqnarray}
\langle n_j(t) \rangle&=
\sum_{n\even}\tilde{h}_{n-j}(t)\tilde{f}_{n-j}(t)-
\sum_{n\odd}\tilde{g}^2_{n-j}(t)\nn
&+\sum_{\substack{n<m \cr \even}}\left[\tilde{g}_{n-j}(t)\tilde{h}_{m-j}(t)
  -\tilde{g}_{m-j}(t)\tilde{h}_{n-j}(t)\right](-1)^\frac{n+m}{2}\nn
&-\sum_{\substack{n<m \cr \odd}}\left[\tilde{g}_{n-j}(t)\tilde{f}_{m-j}(t)
-\tilde{g}_{m-j}(t)\tilde{f}_{n-j}(t)\right](-1)^\frac{n+m}{2}\nn
&+\sum_{\substack{m \odd \cr n \even}}{\rm sgn}(n-m) \tilde{h}_{n-j}(t)\tilde{f}_{m-j}(t) (-1)^\frac{n+m+1}{2}\nn
&-\sum_{\substack{m \even \cr n \odd}} {\rm
  sgn}(n-m)(-1)^\frac{n+m}{2}
\tilde{g}_{n-j}(t)\tilde{g}_{m-j}(t)(-1)^\frac{n+m}{2}\ .
\label{njoft}
 \end{eqnarray}
The time evolution of the particle density \fr{njoft} is shown in
Fig.~\ref{Fig:Neel}. As we are working at balanced particle creation
and annihilation ($J_3=J_4$), $\llangle n_j(t)\rrangle$ relaxes to $1/2$ at late times for all
values of $j$.

\section{Two defect sector}
\label{sec:Defects}
\begin{figure}[ht]
\caption{Eigenvalues of A' (red diamonds), A (blue triangles) for $L=12, J_+=1.0, J_-=0.9$. Central inset is magnified such that both axes run from $\pm 3\times 10^{-5}$.}
\centering
\includegraphics[width=0.7\textwidth]{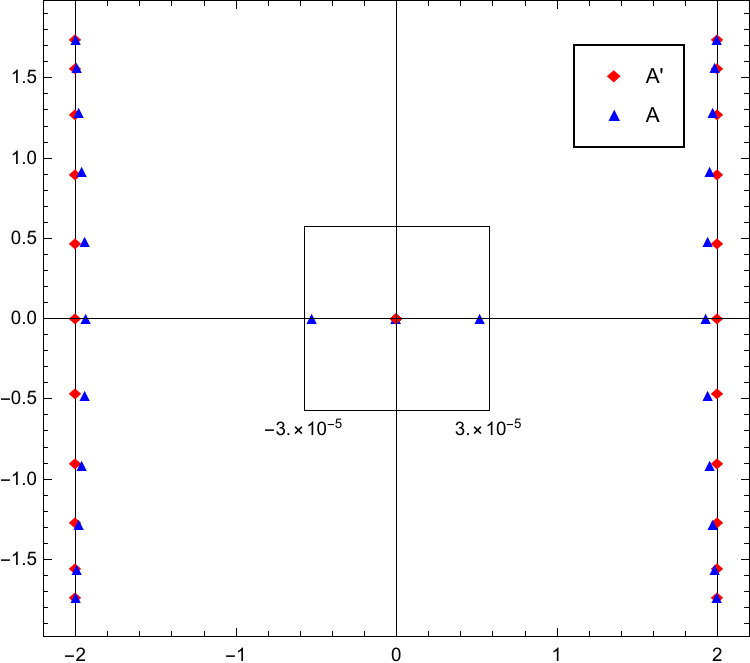}
\label{Fig:DefectSpectrum}
\end{figure}

We now turn to the defect sector problem. The Lindbladian in the
sector with $q$ defects can be written as a sum over quadratic open
chain Lindbladians of the form  
\begin{equation}
{\cal L}_{M}=-J_+(M+1)+ \sum_{j=1}^{M-1} \Big\{ J_2 c^\dagger_j c_{j+1}
    +J_1 c^\dagger_{j+1}c_j -J_+ \left(c^\dagger_{j+1}c^\dagger_j+ c_jc_{j+1}\right) \Big\}.
\end{equation}
As these are not Hermitian the standard analysis of Lieb, Schultz and
Mattis \cite{LSM61} for diagonalizing Hamiltonians quadratic in fermionic
creation/annihilation operators does not apply. We
therefore proceed as in Section \ref{Sec:Imbalanced}, but find it
advantageous to switch to Majorana fermions \cite{prosen2008third}
\be
a_{2j-1}=c_j+c_j^\dagger \  ,\quad
a_{2j}=i(c_j-c_j^\dagger).
\ee
In terms of the Majorana operators ${\cal L}_M$ is expressed as
\be
{\cal L}_M=-(M+1)J_+ +\frac{1}{4}a\cdot A \cdot a,
\ee
where here and elsewhere $(\cdot)$ represents the dot product with no
complex conjugation- that is, $a\cdot A\cdot a = \sum_{ij} a_i A_{ij}
a_j$. $A$ is a $2M\times 2M$ anti-symmetric, block tridiagonal matrix
equal to $A=K\otimes C -K^T\otimes C^T$ where $K_{jk}=\delta_{j,k-1}$
and $C$ is given by
\be
C=\begin{pmatrix}
J_- & -2iJ_+ \cr
0 & J_-   \end{pmatrix}.
\ee
Assuming $A$ to be diagonalizable, anti-symmetry ensures its
eigenvalues come in pairs $\pm\beta_j$ which we order as
$\beta_1,-\beta_1\dots$. We then normalize the eigenvectors according to
\be
\vec{v}_r\cdot\vec{v}_s=(\sigma^x\otimes\id )_{rs}\ .
\label{Eq:NormalisedEvecs}
\ee
In fact, the complex eigenvalues also come in complex conjugate
pairs. This can be seen by noting that one obtains $A^*$ from $A$ by
conjugating $C$ by $\sigma^z$. In particular this means that if
$A\vec{v}=\beta \vec{v}$ then also 
\begin{equation}
    A(\id_n \otimes \sigma^z)\vec{v}^* = \beta^* (\id_n \otimes \sigma^z
    )\vec{v}^*\ .
\end{equation}
Finally we define new fermion operators by
\begin{equation}
b_j=\vec{v}_{2j-1}\cdot{\bf a}/\sqrt{2}\ ,\quad 
b'_j=\vec{v}_{2j}\cdot{\bf a}/\sqrt{2}.
\end{equation}
These fulfil simple anticommutation relations due to \fr{Eq:NormalisedEvecs}
\be
\{b_j,b_k\}=0=\{b'_j,b'_k\}\ ,\quad \{b_j,b'_k\}=\delta_{j,k}\ ,
\ee
and diagonalise the Lindbladian
\begin{equation}
    {\cal L}_M = -(M+1)J_+ +\frac{1}{2} \sum_k \beta_k -\sum_k \beta_k b'_k b_k.
    \label{Eq:Majoranas_Diag}
\end{equation}
For the matrix $A$ in our problem it is not a simple matter to
find a closed form analytic expression for the spectrum, but we can
gain insight into what the solutions look like by deforming our
Lindbladian by adding a boundary term $J_-(n_1-n_L)$. We stress that
the resulting Lindbladian is unphysical. Then the matrix $A$ is
modified to $A'$ 
\be
     A' = \begin{pmatrix}
        B & -C^{\rm T} &0 & \dots & 0 
        \cr C & 0 & -C^{\rm T} & 0 & \vdots 
        \cr 0 & C & 0 & \ddots & 0
        \cr\vdots & 0 & \ddots & 0 & -C^{\rm T}
        \cr 0 & \dots & 0 &  C & -B\end{pmatrix}\ , \quad 
    B = -iJ_-\begin{pmatrix}
        0 & 1 \cr -1 & 0\end{pmatrix}.
    \label{Eq:DefAprime,B}
\ee
It is now straightforward to obtain the eigenvectors of the matrix
$A'$. We make an ansatz $\vec{v}=(v_1,v_2,\dots)^{\rm T}$ where
\be    
v_n = z^n \begin{pmatrix} 1 \cr i z\end{pmatrix} + A_z(-1)^n
        z^{-n}\begin{pmatrix}1 \cr -iz^{-1}
          \end{pmatrix}\ .
    \label{Eq:BdryAnsatz}
\ee
For this to be an eigenvector we require $A_z=z^2$ and $z$ to satisfy
\be
    0=(z^{2M}-(-1)^M)(\Delta z^2-2z-\Delta).
\ee
The associated eigenvalues are then given by
\be
    \lambda_z = 2J_+ +J_-(z^{-1}-z).
\ee
This only gives rise to $M$ linearly independent eigenvectors, all
with non-negative eigenvalues. We however get the full spectrum
using this ansatz by reflecting in the imaginary axis. 
Thus in this case we find that the positive real part eigenvalues
consist of $M-1$ values of $z$ that are roots of unity $z=e^{-ik}$
which recovers the periodic boundary condition result. There are also
two eigenvalues that are exactly $0$ - for our actual boundary
conditions these become two small real eigenvalues $\pm \lambda_0$
(they cannot be complex as the requirement that $\beta^*$ is an
eigenvalue would then give four nearly zero eigenvalues which is too
many). We plot the eigenvalues in the complex plane in  Figure
\ref{Fig:DefectSpectrum} for both $A$ and $A'$ to highlight the impact
of removing the boundary potential. 

\section{Transverse correlation function}
\label{sec:FM}
We now turn to observables that involve defects. We focus on the
particular example of an initial product state with ferromagnetic order along
some direction in spin space
\begin{equation}
    |\psi(0)\rangle = \left(1+|\alpha|^2\right)^{-L/2}\bigotimes_{m=1}^{L} \big[ |\uparrow\rangle + \alpha |\downarrow\rangle \big]_m.
\end{equation}
Our aim is to determine
\be
S_{0,\ell+1}^{+-}=\Tr \left[\rho(t) \sigma^+_L\sigma^-_{\ell+1}\right].
\ee
As we showed above in \fr{Spm2} this involves only the projection of
$|\rho(t)\rangle$ onto the subspace with two defects
\be
S_{0,\ell+1}^{+-}=\llangle\idc|\sigma^+_0\sigma^-_{\ell+1}\Pi_{0,\ell+1}|\rho(t)\rrangle
=\llangle\idc|\sigma^+_0\sigma^-_{\ell+1}e^{{\cal L} t}\Pi_{0,\ell+1}|\rho(0)\rrangle,
\ee
where
\be
\Pi_{0,\ell+1}=E_0^{33}\otimes_{j=1}^{\ell}P_j E_{\ell+1}^{22}\otimes_{k=\ell+2}^{L-1}P_k\ .
\ee
Applying $\Pi_{0,\ell+1}$ to the initial density matrix $|\rho(0)\rrangle
=|\psi(0)\rangle \langle \psi(0)|$ gives 
\begin{equation}
\Pi_{0,\ell+1} |\rho(0)\rrangle = (1+|\alpha|^2)^{-L} |\alpha|^2 |3\rrangle_L \otimes |\rho_{[1,\ell]}\rrangle \otimes |2\rrangle_{\ell+1} \otimes |\rho_{[\ell+2,L-1]}\rrangle,
\end{equation}
where
\be
|\rho_{[a,b]}\rrangle\equiv\otimes_{m=a}^{m=b}\left[|1\rrangle
  +|\alpha|^2|4\rrangle\right]_m.
\ee
We now see that the transverse spin-spin correlation function reduces in this initial state to
\begin{align}
\Tr \left[\rho(t) \sigma^+_1\sigma^-_\ell\right]&=
\gamma (1+\gamma)^{-L} \tilde{G}_{\ell}(t) \tilde{G}_{L-\ell-2}(t) \nn
\tilde{G}_{N}(t) &= \llangle \id_{[1,N]} | e^{{\cal L}_N t} | \rho_{[1,N]} \rrangle  .
\label{Eq:FMCorrelationFn}
\end{align}
Here we have defined $\gamma=|\alpha|^2$ since all quantities depend
only on $\gamma$ and have separated out an overall factor
$\gamma(1+\gamma)^N$ for convenience.

\begin{figure}[h]
\caption{Full correlation function $S^{+-}_{0,2}(t)$ for $L=30$ sites,
  $J_-=0.9, J_+=1.0$ and $\gamma=0.9$.}  
\centering
\includegraphics[width=0.75\textwidth]{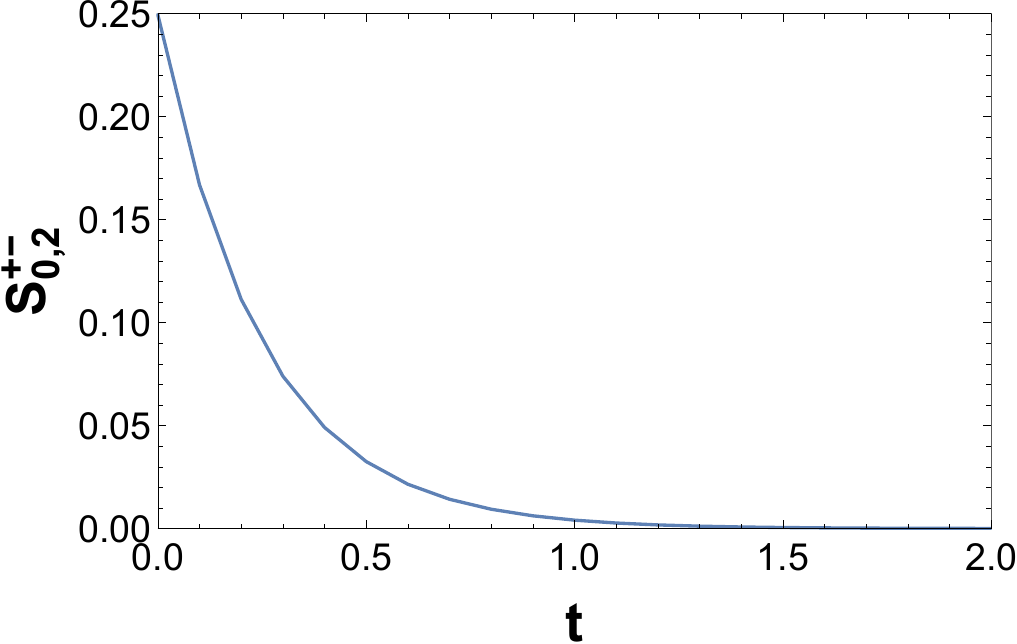}
\label{Fig:FullCorrelator}
\end{figure}

\begin{figure}[ht]
(a)\includegraphics[width=.4\textwidth]{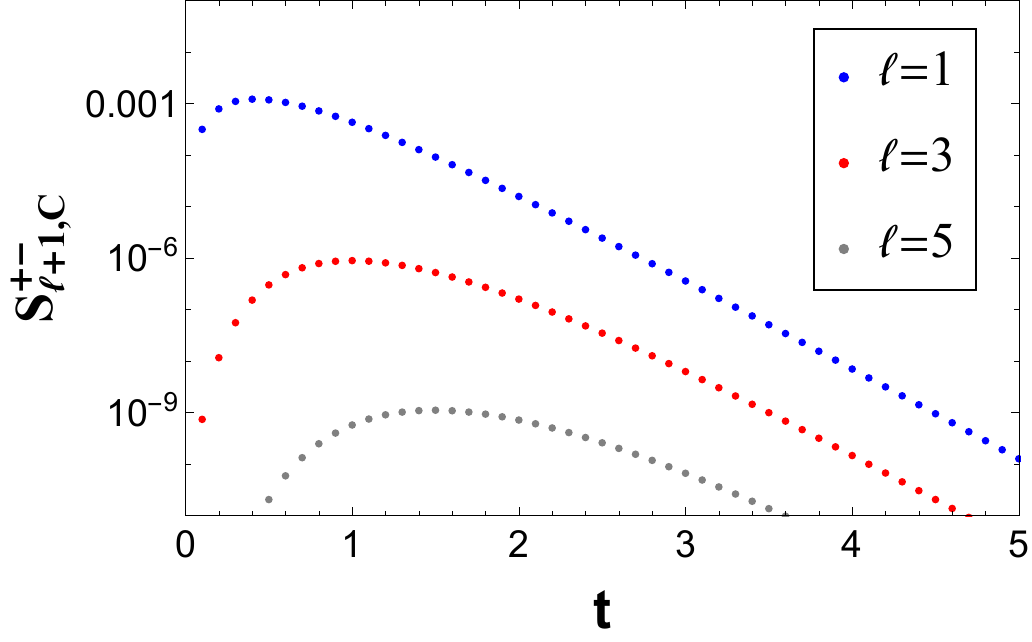}\qquad
(b)\includegraphics[width=.4\textwidth]{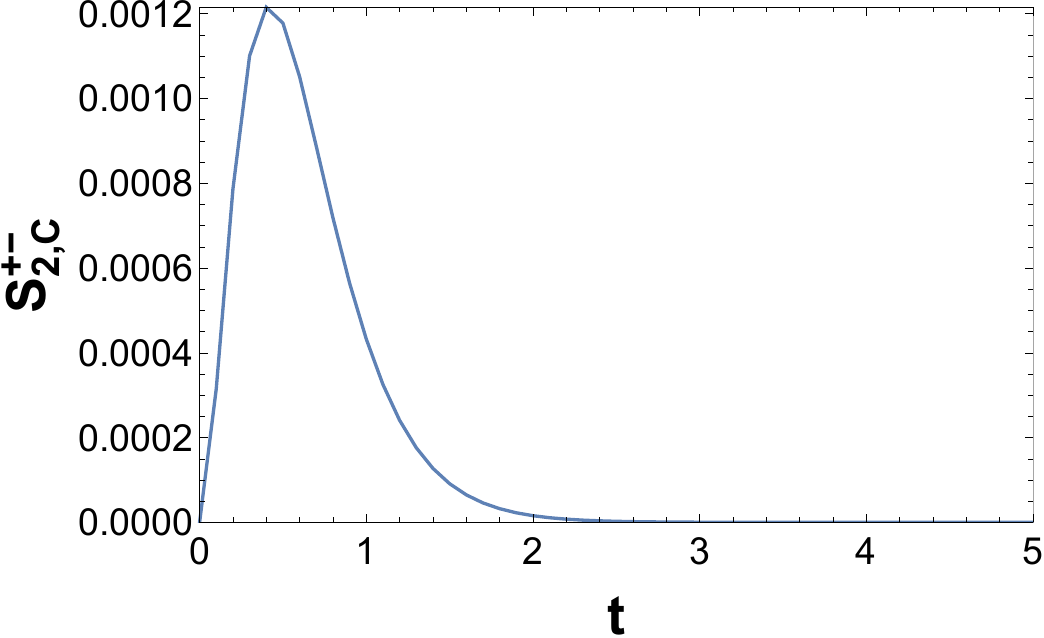}
\caption{Connected correlation function for chain length $L=30$ and
  $J_-=0.9, J_+-1.0, \gamma=0.9$. (a) Connected correlations decay
  exponentially with separation $d=\ell+1$. (b) Connected correlation
function for $\ell=1$, showing correlation generation.}
\label{Fig:ConnectedCorrelations}
\end{figure}

\begin{figure}
(a)\includegraphics[width=.4\textwidth]{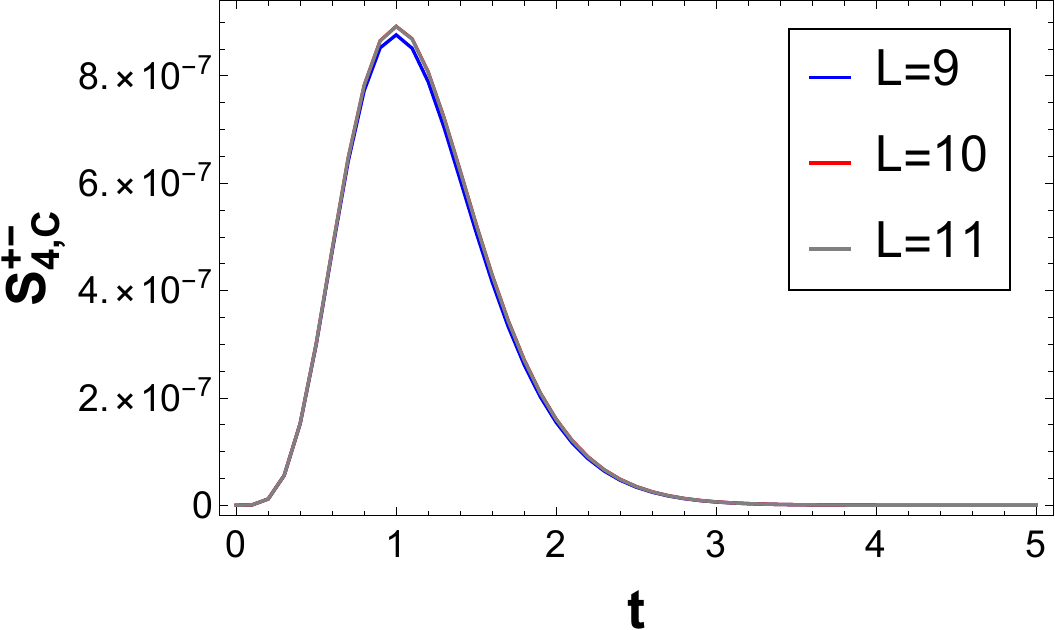}\qquad
(b) \includegraphics[width=.4\textwidth]{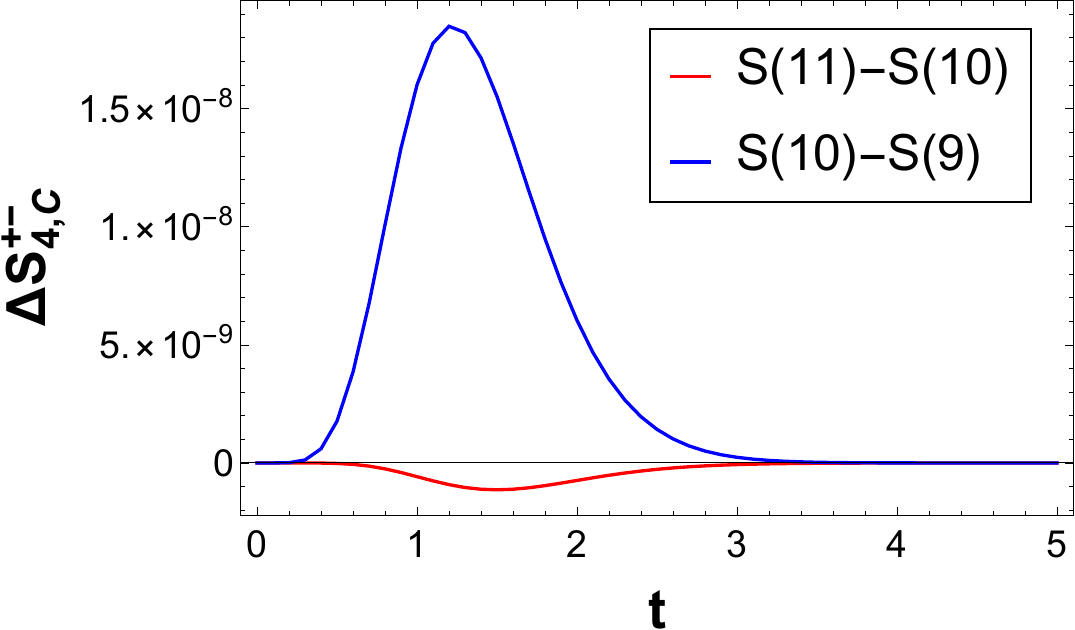}
\caption{Finite size effects for $\ell=3$. $J_-=0.9, J_+-1.0,
  \gamma=0.9$: (a) Correlation function for three different chain
  lengths $L=8,9,10$. (b) Residuals between correlation function at
  $L=9,10$ and $L=10,11$. } 
\label{Fig:ThermodynamicLimit}
\end{figure}

The propagators $\tilde{G}$ are defined on the finite chains discussed
in Section \ref{sec:Defects} and can be expressed in terms of fermions as
\be
\tilde{G}_{N}(t)=
{\phantom{\Bigrrangle}}_0\Bigllangle 0\Big|\prod_{j=0}^{N-1}\big(1+c_{N-j}\big) e^{{\cal
    L}_N t}\prod_{k=1}^N\big(1+\gamma
c^\dagger_k\big)\Big|0\Bigrrangle_0\ .
\ee
As shown in \ref{App:Identities} we can rewrite this as
\be
\tilde{G}_N(t)=
{}_0\llangle 0|(1+X)e^{Y} e^{{\cal
    L}_Nt}e^{\gamma^2Y^\dagger}(1+\gamma X^\dagger)|0\rrangle_0\ , \label{Eq:GTilde}
\ee
where
\be
X= \sum_{j=1}^N c_j\ , \quad 
Y= \sum_{n<m}c_mc_n\ .
\ee
Using fermion parity conservation this simplifies to
\begin{equation}
\tilde{G}_N(t)=
{}_0\llangle 0|e^{Y} e^{{\cal
    L}_{N}t}e^{\gamma^2Y^\dagger}|0\rrangle_0 
+\gamma\ {}_0\llangle 0|Xe^{Y} e^{{\cal
    L}_{N}t}e^{\gamma^2Y^\dagger}X^\dagger|0\rrangle_0\ .
\label{g1ell0}
\end{equation}
The two terms above can be written in the form
\be
\tilde{G}^{(\alpha)}_N=\Tr \left[\tilde{\rho}^{(\alpha)} e^{Y} e^{{\cal
    L}_{N}t}e^{\gamma^2Y^\dagger}\right],\quad\alpha=1,2\ ,
\label{Eq:GTraces}
\ee
where $e^Y$,$e^{{\cal L}_{N}t}$, $e^{\gamma^2Y^\dagger}$ are
all manifestly Gaussian as are the $\tilde{\rho}^{(\alpha)}$ since
they are the ground states of the quadratic Hamiltonians
\be
H_1 = \sum_{j=1}^N n_j\ , \quad H_2=-n(p=0)+\sum_{p\neq 0}n(p)\ .
\ee
Thus \fr{Eq:GTraces} is now in the form of the trace of a product of
Gaussian operators and can be evaluated. The procedure for this is
given in detail in \ref{App:Identities}. Here we outline the two key
steps to the evaluation. The first step is to realise that since a
product of Gaussian operators is Gaussian, we have
\be
\rho^{(\alpha)} = \frac{e^{\gamma^2
Y^\dag}\tilde{\rho}^{(\alpha)}e^Y}{Z^{(\alpha )}}=\frac{1}{\mathcal{Z}(W^{(\alpha)})}e^{a\cdot
W^{(\alpha)}\cdot a/4}. 
\ee
Here, $Z^{(\alpha)}$ and ${\cal Z}(W^{(\alpha)})$ are two different normalisation factors defined each defined such that $\Tr \rho^{(a)}=1$. The $Z^{(a)}$ are calculated in \ref{App:Gamma} and given by \ref{Eq:Z}. Writing the time evolution operator in the form
\be
e^{{\cal L}_Nt}=e^{\frac{1}{4}a\cdot A_N \cdot a}\ ,
\ee
we then obtain the following expression for the propagators ({\emph
  cf.} \ref{App:Identities} )
\be
\Tr \left[\rho^{(\alpha)} e^{\frac{1}{4}a\cdot A_N\cdot
    a}\right]=Z^{(\alpha)} \left( \frac{{\rm 
      det}(e^{W^{(\alpha)}}e^{A_N}+e^{-A_N}e^{-W^{(\alpha)}}+2)}{{\rm
      det}(e^{W^{(\alpha)}}+e^{-W^{(\alpha)}}+2)}\right)^{1/4} \ .
    \label{Eq:GaussianExpectation}
\ee
The second step is to use the fact that a Gaussian is determined by
its second moments to change from working with the density matrix
$e^{a\cdot W\cdot a/4}$ itself to instead working with its correlation
matrix $\Gamma_{mn}=\Tr[\rho a_n a_m]-\delta_{mn}$. We calculate the
latter in \ref{App:Gamma} by rewriting the trace as an inner
product which can be evaluated in terms of Jordan-Wigner spins. Once
$\Gamma$ is found $W$ is obtained through $\Gamma=\tanh\frac{W}{2}$,
or equivalently 
\be
    e^W = (1-\Gamma)^{-1}(1+\Gamma).
\ee
This then leads to an apparent difficulty since the correlation matrices
corresponding to $\rho^{(\alpha)}$ (which are fixed through our choice
of initial condition) satisfy $(\Gamma^{(\alpha)})^2=1$, implying that they have only eigenvalues equal to $\pm 1$ and the
method set out above appears to break down. This issue can be dealt
with by noting that
\be
\Gamma^{(\alpha)}= \Pi^{(\alpha)}_+ - \Pi^{(\alpha)}_-\ ,
\ee
where $\Pi^{(\alpha)}_{\pm}$ are projectors onto the two $N$
dimensional subspaces corresponding to eigenvalues $1$ and $-1$ of
$\Gamma^{(\alpha)}$ respectively. These would correspond to
eigenvalues $\pm\infty$ in $W$, which we regulate by setting them
equal to $\pm \Lambda$ and taking the limit $\Lambda\to\infty$ in the
end of the calculation. That is, we put: 
\be
    e^{W^{(\alpha)}/2}=e^{\Lambda/2}\Pi^{(\alpha)}_+ +
    e^{-\Lambda/2}\Pi^{(\alpha)}_-\ ,
\ee
which simplifies \fr{Eq:GaussianExpectation} to read
\begin{align}
\Tr \left[\rho^{(\alpha)} e^{\frac{1}{4}a\cdot A_Nt \cdot a}\right]
    &= \left( {\rm det}(\Pi^{(\alpha)}_+ e^{A_Nt}+\Pi^{(\alpha)}_-
  e^{-A_Nt})\right)^{1/4} \equiv d^{(\alpha)}_N \ .
    \label{Eq:DetProjectors}
\end{align}
This yields a simple expression for the propagator
\begin{eqnarray}
\tilde{G}_N(t) =& e^{-J_+(N+1)t}\Big[Z^{(1)} d^{(1)}_N
  + \gamma Z^{(2)}d^{(2)}_N
  \Big].
    \label{Eq:FM_G}
\end{eqnarray}
Substituting this into \fr{Eq:FMCorrelationFn} then gives the
transverse correlation function
\begin{eqnarray}
    S^{+-}_{0,\ell+1}(t) = \frac{\gamma}{(1+\gamma)^L} & e^{-J_+Lt} \left(Z^{(1)} d_{\ell}^{(1)}+\gamma Z^{(2)} d_{\ell}^{(2)}\right)  \nn
    &\times \left( Z^{(1)}d_{L-\ell-2}^{(1)}+\gamma Z^{(2)} d_{L-\ell-2}^{(2)} \right).
    \label{Eq:Full2PtFn}
\end{eqnarray}
The determinants $d^{(\alpha)}_N$ can now be straightforwardly
computed numerically. In Fig.~\ref{Fig:FullCorrelator} we plot the
transverse correlator at separation $2$, $S^{+-}_{0,2}(t)$, as a
function of time.
We observe that the correlator decays quite quickly, and
monotonically, from its initial value. We note however that this is
the full correlation function and that more physically interesting is
the connected correlator  
\begin{equation}
    S^{+-}_{\ell+1,C} = \Tr [S^+_0 S^-_{\ell+1} \rho(t)]-\Tr [S^+_0\rho(t)]\Tr [S^-_{\ell+1} \rho(t)].
\end{equation}
Here we have use the translation invariance of our initial condition to express the correlation function in terms of only the distance between the defects (note that this is $\ell+1$ and not $\ell$). The one point functions depend on the same propagators as the 2-point functions since
\begin{equation}
     {\rm Tr}[S^+_0 \rho(t)]= \llangle \idc | E^{13}_0 E^{24}_0 |\rho(t)\rrangle = \frac{\alpha}{1+\gamma} G_{L-1},
\end{equation}
which gives
\begin{equation}
      S^{+-}_{\ell+1,C}=\frac{\gamma}{(1+\gamma)^2}\left[ G_{\ell}G_{L-\ell-2}-G_{L-1}^2 \right].
\end{equation}
Where we have expressed this in terms of
$G_N=(1+\gamma)^{-N}\tilde{G}_N$ as this is more natural. In
particular, since our initial state was a product state we have
$G_N(0)=1$ for all $N$ and so the connected correlation function is
initially $0$, indicating no correlations. We then expect that the
Lindblad evolution will correlate neighbouring sites. This is
countered by the fact that the steady state values of observables are
all governed by the diagonal subspace values and so the connected
correlations must go to zero at long times. In Figure
\ref{Fig:ConnectedCorrelations}(a) we plot the connected correlation
between sites $1$ and $3$. We are able to observe that the dissipative
dynamics does produce some correlations although they are small. Given
that they also exponentially decay, the correlation generation would
most likely not be visible had we started in an initially correlated
state. In Figure \ref{Fig:ConnectedCorrelations}(b) we plot the
corresponding values for varying site separations and note an
approximately exponential decrease with distance. We perform these
calculations for total chain lengths of $L=30$, one might wonder if
this is large enough to be essentially in the thermodynamic limit (in
the sense that finite size effects are small enough to neglect). In
fact, we find that the numerical values of the connected correlator
vary very little as we increase $L$ so long as it is larger than twice
the separation $\ell+1$. To show this we plot the connected
correlator for $\ell=3$ for $L=8,9,10$ in Figure
\ref{Fig:ThermodynamicLimit}(a). Since the difference between the
result for $9,10$ is too small to be visible, we plot the residual
(along with the corresponding residual for $L=10,11$) in Figure
\ref{Fig:ThermodynamicLimit}(b). 

\section{Conclusions}
\label{Sec:Conclusion}
We have considered a dissipative many-particle quantum system
described by a Lindblad equation that for particular initial conditions
reduces to an asymmetric simple exclusion process with additional
pair creation and annihilation terms. The Lindbladian exhibits
operator-space fragmentation and for particular pair
creation/annihilation rates the model can be mapped to free 
fermions. The model thus extends the class of solvable Lindblad 
systems and in particular provides a concrete example of a setting
where operator-space fragmentation can be used to compute correlation
functions exactly. We have restricted attention to initial product
states in order to make calculations simpler as well as to allow us to
see the generation of correlations through dissipation in our
model. Even though the initial states we have considered here are quite
simple the analysis is not straightforward. It would be interesting to
attempt to generalize our analysis to the case of entangled initial
states. It would also be interesting to study the time evolution of
entanglement measures such as entanglement negativity within this
model. 

\ack 
We thank Denis Bernard and Lorenzo Piroli for helpful
discussions and Aleksandra Ziolkowska for collaboration during the
early stages of this work. This work was supported in part by the
EPSRC under grant EP/S020527/1.

\appendix
\section{Fermion identities}
\label{App:Identities}
\subsection{Mixed parity fermion products}

To arrive at Eq \fr{Eq:GTilde} the key identity is that for any
collection of mutually anti-commuting variables $\{\zeta_i\}_{i=1}^N$
the following holds
\be
    \prod_{i=1}^{N}(1+\zeta_i)=\left(1+\sum_{i=1}^N \zeta_i\right)\prod_{1\leq j<k\leq N}(1+\zeta_j \zeta_k).
    \label{Eq:MixedParityIdentity}
\ee
This identity immediately provides a convenient decomposition into even and odd
fermion parity parts. It can be proven by focussing on the odd and
even components and using induction. To do so note that the even terms
have the form 
\be
\mathbf{E}\left[\prod_{i=1}^N(1+\zeta_i)\right] = \sum_{\ontop{k\text{
      even}}{k\leq N}}\left( \sum_{1\leq i_1 < \dots < i_k \leq N}\zeta_{i_1}\dots \zeta_{i_k} \right).
\ee
The counterpart for the odd terms is completely analogous. When
multiplying this by $\sum_{j=1}^N \zeta_j$ the result will be a sum of
$(N-k)\binom{N}{k}$ non-zero terms, each of which contain $k+1$
distinct $\zeta$'s. Moreover, in each term $k$ of the $\zeta$'s will
be in ascending order by construction, with the final one appearing in
each possible position. Thus $k/2$ pairs will cancel and the remaining
$\frac{(N-k)}{k+1}\binom{N}{k}=(N-k-1)\binom{N}{k+1}$ terms are
precisely those in the corresponding expansion of the odd part. We
thus need only prove by induction the statement about the even terms
\be
    \sum_{k=2m}\left( \sum_{1\leq i_1 < \dots < i_k \leq
      N}\zeta_{i_1}\dots \zeta_{i_k} \right) = \prod_{1\leq j<k\leq
      N}(1+\zeta_j\zeta_k)\ .
\ee
To this end we assume the induction hypothesis up to $N-1$ and notice that for $N$ sites we can rewrite the product of quadratic terms as
\be
        \prod_{i < j}^{N-1}(1+\zeta_i\zeta_j)\prod_{\omega=1}^{N-1} (1+\zeta_{\omega}\zeta_N)=\prod_{i<j}^{N-1}(1+\zeta_i\zeta_j)(1+\sum_{\omega=1}^{N-1} \zeta_\omega \zeta_N).
\ee
We then use the induction hypothesis on the first factor on the right
hand side. When multiplied by the second factor two things can happen:
(i) it gets multiplied by 1, thus generating all possible even terms
not including $\zeta_N$, (ii) it gets multiplied by $(\sum_{\omega=1}^{N-1}
\zeta_\omega) \zeta_N$. For the latter note that multiplying by
$\sum \zeta_\omega$ generates all possible odd expressions without $\zeta_N$
and multiplying by $\zeta_N$ at the end then gives the desired
result. Along with the observation that the base case of $N=0$ is
trivial this completes the proof of \fr{Eq:MixedParityIdentity}.  

In the context of Eq \fr{Eq:GTilde} we set $\zeta_i=\gamma c_i^\dag$
so that we have
\be
\prod_{k=1}^N(1+\gamma c_k^\dag) =\prod_{1\leq m<n\leq N}(1+\gamma^2
c_m^\dag c_n^\dag)(1+\gamma \sum_{k=1}^N c_k^\dag) \ .
\ee
This is the desired simplification upon defining $X,Y$ in the main
text and applying the standard result that
$e^{Y^\dag}=\prod_{m<n}(1+c^\dag_m c^\dag_n)$ where
$Y^\dag=\sum_{m<n}c_m^\dag c_n^\dag$. 
\subsection{Trace of Gaussian operators}
The main result required in the derivation of
\fr{Eq:GaussianExpectation} is the identity \cite{Fagotti2010}
\begin{equation}
    \Tr[e^{a\cdot W\cdot a/4}] = \sqrt{{\rm det}(e^{W/2}+e^{-W/2})}.
    \label{Eq:TraceToDet}
\end{equation}
Here $a_j$ are Majorana fermions and $W$ an antisymmetric matrix. We
note that \fr{Eq:TraceToDet} is easy to establish if $W$ is
diagonalizable. In that case its eigenvalues come in pairs $\pm\beta_k$
and the left hand side becomes
\begin{equation}
\Tr [e^{\sum_k \beta_k(\frac{1}{2}-n_k)}]=\prod_{\text{Re}\beta_k>0}\left[e^{\beta_k/2}+e^{-\beta_k/2}\right].
    \label{Eq:TraceToDetProof}
\end{equation}
Because $W$ is anti-symmetric, its eigenvalues come in $\pm$ pairs and
so the determinant contains exactly two copies of each factor on the
right hand side of \fr{Eq:TraceToDetProof}. This then establishes
\fr{Eq:TraceToDet}. If $W$ is not diagonalizable then we define 
\be 
    f[W]=\Tr[e^{a\cdot W\cdot a/4}]-\sqrt{\det(e^{W/2}+e^{-W/2})}.
\ee
Since $f[W]=0$ for all diagonalisable matrices (which is a dense subset of all matrices) and $f$ is a continuous function, we have $f=0$ identically.

We also make heavy use of a result following from the
Baker-Campbell-Hausdorff formula, namely that for Majorana fermions
$a_j$ normalised such that $\{a_i,a_j\}=2\delta_{ij}$  
\be
    e^{\frac{1}{4}a\cdot W_1 \cdot a}e^{\frac{1}{4}a\cdot W_2 \cdot
      a}=e^{\frac{1}{4}a\cdot W_3 \cdot a}\ ,
\ee
where $e^{W_3}=e^{W_1}e^{W_2}$. Along with \fr{Eq:TraceToDet} this
allows us to write 

\bea
\Tr[e^{\frac{1}{4}a\cdot W_1\cdot a}e^{\frac{1}{4}a\cdot W_2\cdot a}]&=
\sqrt{{\rm det}(e^{W_3/2}+e^{-W_3/2})} \nn
&= \big[{\rm det}(e^{W_3}+e^{-W_3}+2)\big]^{1/4}\ .
\eea

\section{Correlation matrices}
\label{App:Gamma}

The two correlation matrices we need are given by the inner products
\begin{align}
    \Gamma^{(1)}_{mn}+\delta_{mn} &= \frac{\llangle 0 | e^Y a_n a_m
      e^{\gamma^2 Y^\dag} | 0 \rrangle}{\llangle 0 |e^Y e^{\gamma^2
        Y^\dag}|0\rrangle }\ ,\\
    \Gamma^{(2)}_{mn}+\delta_{mn} &= \frac{\llangle 0 | X e^Y a_n a_m e^{\gamma^2Y^\dag}X^\dag |0\rrangle}{\llangle 0 | X e^Y e^{\gamma^2Y^\dag} X^\dag |0\rrangle}.
\end{align}
The denominators are equal to the normalization factors
$Z^{(1)},Z^{(2)}$ that appear in the final result
\fr{Eq:Full2PtFn}. Both numerators and denominators can be found by
making use of \fr{Eq:MixedParityIdentity} in the form 
\begin{align}
\prod_{j=1}^N \left(1+\gamma c_k^\dag\right) + \prod_{j=1}^N
\left(1-\gamma c_k^\dag\right) =& 2e^{\gamma^2 Y^\dag}\ ,
\\    \prod_{j=1}^N \left(1+\gamma c_k^\dag\right) - \prod_{j=1}^N
\left(1-\gamma c_k^\dag\right) =& 2\gamma Xe^{\gamma^2 Y^\dag} \ .
\label{prods}
\end{align}
Using \fr{prods} we can express the correlation matrices as
\begin{eqnarray}
    \Gamma^{(1)}_{mn}+\delta_{mn} &=
    \frac{g^{++}_{mn}+g^{+-}_{mn}}{Z^{++}+Z^{+-}}\ , \quad
    \Gamma^{(2)}_{mn}+\delta_{mn} =
    \frac{g^{++}_{mn}-g^{+-}_{mn}}{Z^{++}-Z^{+-}}\ ,
\end{eqnarray}
where we have defined
\begin{eqnarray}
    g^{\sigma\sigma'}_{mn}=&\llangle 0 |\prod_{j=0}^{N-1}(1+\sigma c_{N-j})a_n a_m
    \prod_{k=1}^N(1+\sigma' \gamma c_k^\dag) |0\rrangle\ ,
    \label{Eq:gplusminus} \\
    Z^{\sigma\sigma'} =& \llangle 0 |\prod_{j=0}^{N-1}(1+\sigma c_{N-j})
    \prod_{k=1}^N(1+\sigma' \gamma c_k^\dag) |0\rrangle\ .
\end{eqnarray}
A simple calculation then gives that $g_{mn}^{++}=g_{mn}^{--}$ and
$g_{mn}^{+-}=g_{mn}^{-+}$ and likewise for $Z^{\sigma\sigma'}$.  
Explicit expressions for $g_{mn}^{\sigma\sigma'}$ and $Z^{\sigma\sigma'}$
are readily obtained by reverting to their respective representations
in terms of spins (i.e. undoing the Jordan-Wigner transformation). We find
\bea
    Z^{(1)}= \frac{1}{2}\left[(1+\gamma)^N+(1-\gamma)^N\right]\ , \nn
    Z^{(2)}= \frac{1}{2}\left[(1+\gamma)^N-(1-\gamma)^N\right],
    \label{Eq:Z}
\eea
and $\Gamma^{(\alpha)}$ are anti-symmetric $2N\times 2N$ block
matrices of the form
\begin{equation}
\Gamma^{(\alpha)} =\frac{1}{x^N -(-1)^\alpha y^N}\begin{pmatrix}
\Gamma_0^{(\alpha)} & - (\Gamma^{(\alpha)}_1)^{\rm T} &\dots
&-(\Gamma^{(\alpha)}_{N-1})^{\rm T} \cr
\Gamma_1^{(\alpha)} & \Gamma_0^{(\alpha)}&\dots &\vdots  \cr
                    \vdots & \dots& \dots & \vdots \cr
                    \Gamma_{N-1}^{(\alpha)}& \dots &\dots & \Gamma_0^{(\alpha)}
                    \end{pmatrix}\ ,\quad \alpha=1,2.
\end{equation}
The $2\times 2$ blocks are given by
\begin{align}
  \Gamma^{(\alpha)}_0 &= 
  \begin{pmatrix}0& i(f_{N-1} \pm
      f_1)\cr-i(f_{N-1} -(-1)^\alpha f_1)& 0\end{pmatrix}\ ,      \nn
\Gamma^{(\alpha)}_a &= \begin{pmatrix} f_{N-a} -(-1)^\alpha
      f_{a} & -i(f_{a-1} -(-1)^\alpha f_{N-a+1}) \cr -i(f_{a+1} -(-1)^\alpha f_{N-a-1})
      & -(f_{N-a} -(-1)^\alpha f_{a})
      \end{pmatrix}\ ,
\end{align}
where we have defined
\be
x=1+\gamma\ ,\quad y=1-\gamma\ ,\quad f_a =x^{a}y^{N-a}\ .
\ee
One can verify that $(\Gamma^{(\alpha)})^2=\id$. Since
$\Gamma^{(\alpha)}$ is anti-symmetric it therefore has equal numbers
of eigenvalues $\pm 1$, which we used in the main text. The 
eigenvectors depend on $x,y$ and we find these numerically to
determine the correct projectors to use.

\vskip 1cm

\bibliographystyle{utphys2}
\bibliography{bibliography}

\end{document}